# Biofilms as self-shaping growing nematics

Japinder Nijjer[1], Mrityunjay Kothari[2,3], Changhao Li[4], Thomas Henzel[2], Qiuting Zhang[1], Jung-Shen B. Tai[1], Shuang Zhou[5], Sulin Zhang[4,6], Tal Cohen*[2,7], Jing Yan[1,8]*

[1]Department of Molecular, Cellular and Developmental Biology, Yale University, New Haven, CT, USA.

[2]Department of Civil and Environmental Engineering, Massachusetts Institute of Technology, Cambridge, MA, USA.

[3]Department of Mechanical Engineering, University of New Hampshire, Durham, NH, USA

[4]Department of Engineering Science and Mechanics, Pennsylvania State University, University Park, PA, USA.

[5]Department of Physics, University of Massachusetts Amherst, Amherst, MA, USA.

[6]Department of Biomedical Engineering, Pennsylvania State University, University Park, PA, USA.

[7]Department of Mechanical Engineering, Massachusetts Institute of Technology, Cambridge, MA, USA.

[8]Quantitative Biology Institute, Yale University, New Haven, CT, USA.

**Abstract:**

Active nematics are the nonequilibrium analog of passive liquid crystals in which anisotropic units consume free energy to drive emergent behavior. Similar to liquid crystal (LC) molecules in displays, ordering and dynamics in active nematics are sensitive to boundary conditions; however, unlike passive liquid crystals, active nematics, such as those composed of living matter, have the potential to regulate their boundaries through self-generated stresses. Here, using bacterial biofilms confined by a hydrogel as a model system, we show how a three-dimensional, living nematic can actively shape itself and its boundary in order to regulate its internal architecture through growth-induced stresses. We show that biofilms exhibit a sharp transition in shape from *domes* to *lenses* upon changing environmental stiffness or cell-substrate friction, which is explained by a theoretical model considering the competition between confinement and interfacial forces. The growth mode defines the progression of the boundary, which in turn determines the trajectories and spatial distribution of cell lineages. We further demonstrate that the evolving boundary defines the orientational ordering of cells and the emergence of topological defects in the interior of the biofilm. Our findings reveal novel self-organization phenomena in confined active matter and provide strategies for guiding the development of programmed microbial consortia with emergent material properties.

**Main Text:**

Active nematics are collections of anisotropic particles which metabolize free energy to generate mechanical work. Unlike conventional liquid crystals (LCs), they exist far from equilibrium and activity plays an important role in shaping their collective structure and dynamics[1–6]. One prototypical example of active nematics, with non-conserving particle number, are growing colonies of bacterial cells with elongated shapes[7–12]. When bacteria collectively secrete extracellular matrix to adhere to each other and to a substrate, they form multicellular communities known as biofilms[13]. Biofilms grow in diverse set of environments including in the ocean, in soil, and in humans, and as they develop, they take on a rich variety of three-dimensional (3D) morphologies and internal architechtures[14–18]. Moreover, the anisotropic shape of bacterial cells can lead to parallel alignment and nontrivial global organization[19–22], which allows one to use biofilms as model living nematic systems to probe the feedback between evolving boundaries and internal ordering. Understanding this feedback could allow for controlled growth of beneficial biofilms, elimination of harmful ones, and the potential development of a new class of growing active materials that not only respond to but also actively alter their geometry to maximize functionalities.

**Competition between confinement and interfacial forces controls biofilm morphogenesis**

Here, we use confined *Vibrio cholerae* biofilms as the model system to demonstrate the self-shaping and self-organizing capability of a 3D growing nematic system. To focus on the cell organization and mechanical aspect of biofilm growth, we used a locked biofilm-forming strain, labeled as WT*[14,23]. To tune the effect of the boundary, we employed a geometry where the biofilm-forming bacteria were confined between a soft hydrogel and a stiff glass substrate[14]. We varied the stiffness of the overlaying gel by varying the agarose concentration from 0.2% to 2.5%, resulting in shear moduli that ranged from 150 Pa – 150 kPa (Extended Data Fig. 1). The gel mesh size was smaller than the cells and therefore confined them, but large enough to allow free diffusion of nutrient and waste molecules. In each case, the biofilms grew clonally from a single cell into a mature biofilm consisting of thousands of cells. Using time-lapse 3D imaging and cell-segmentation algorithms[14,20], we extracted and tracked the evolution of biofilm architectures at *single-cell* resolution. Figure 1a shows a series of segmented biofilms grown under gels of different concentrations consisting of approximately 8600 cells. We found that as the biofilms matured they developed into one of two bulk shapes indicating two distinct growth modes: under soft confinement ($c \leq 1\%$), the biofilms grew as hemispherical structures, which we label "domes," whereas under stiff confinement ($c \geq 2\%$), the biofilms grew as flatter structures, which we label "lenses" (Fig. 1b). At intermediate gel concentrations ($1\% < c < 2\%$), we observed the coexistence of both lenses and domes. To quantify this shape transition, we measured the contact angle ($\psi$) that the biofilms made on the glass substrate for hundreds of mature biofilms for each condition (Fig. 1c; Extended Data Fig. 2 and 3). Interestingly, $\psi$ exhibited a bifurcation-like transition with increasing stiffness. Biofilms possessed larger $\psi$ when grown under soft gels (median $\psi$ range 101°-121° for $c = 0.2\%$-1%) and smaller $\psi$ when grown under stiff gels (median $\psi$ range 23°-39° for $c = 2\%$-2.5%); at intermediate concentrations ($c = 1.3\%$-1.75%), a bimodal distribution of $\psi$ emerged, with each peak coinciding with either the large $\psi$ (small stiffness, median $\psi$ range 127°-131° for the subset $\psi \geq 75°$) or small $\psi$ (large stiffness, median $\psi$ range 33°-40° for the subset $\psi < 75°$) behavior (Fig. 1d).

In addition, the kinetics of shape evolution also significantly differed between the two regimes, as quantified by the evolution of the maximum height $h_{max}$ and maximum radius $r_{max}$ of the biofilms (Fig. 1e). For soft confinement, the biofilms grew nearly isotropically with the maximum height and radius scaling as $h_{max} \sim N^{1/3}$ and $r_{max} \sim N^{1/3}$, respectively (where $N$ is the number of cells). For stiff confinement, the biofilms grew faster horizontally than vertically, leading to an increasingly anisotropic shape over time. This was reflected in the different scaling laws for biofilm radius and height where $h_{max} \sim N^{1/5}$ and $r_{max} \sim N^{2/5}$, reminiscent of those observed during hydraulic fracturing[24,25]. Correspondingly, we observed two diverging trajectories of $\psi$ (Fig. 1f) where $\psi$ either increased or decreased above ~100 cells.

Previous work suggests cell-substrate friction as a key determinant in biofilm morphogenesis[16,26], which in *V. cholerae* is mainly achieved by two redundant adhesion proteins RbmC and Bap1[27,28]. Upon deleting these adhesins, we found that the critical stiffness at which the shape transition occurred decreased (Extended Data Fig. 4). To further demonstrate the effect of cell-substrate friction on biofilm shape, we generated a strain with an arabinose-inducible expression vector with titratable expression of *bap1*. Indeed, as *bap1* expression increased, the critical stiffness at which the biofilms transitioned from domes to lenses also increased (Fig. 2a). A bimodal distribution of shapes was again observed near the phase boundary in the two-dimensional phase diagram.

To confirm the important role that cell-substrate interactions play in shaping the final biofilm shape, we employed experimentally benchmarked agent-based simulations (ABSs)[14,29]. In the simulations, we introduced a frictional force that resisted the growth-induced sliding of cells parallel to the substrate to mimic the effect of the two adhesins in the experiment. By varying the surface friction coefficient and gel stiffness in the ABSs, we reproduced a similar transition from large to small $\psi$ upon decreasing friction or increasing gel stiffness (Fig. 2b), which suggests that the adhesion proteins indeed control biofilm morphology by increasing the friction between the biofilms and the substrate.

**An energetic model of biofilm morphogenesis explains the shape transition**

To elucidate the origins of the two different growth regimes, we consider the energetics of biofilm growth confined at the bonded interface between a semi-infinite elastic material and a rigid substrate, while accounting for the frictional losses that are experienced by the biofilm as it slides along the substrate. Here we model the biofilm as a volumetrically expanding liquid because it can continuously reorganize itself during growth[20,30,31]. As the biofilm expands, it can either deform the surrounding gel or delaminate the gel from the glass substrate, or both. We consider the total potential energy of the system $U = U_d + U_e$ as the sum of[31]: (1) adhesion energy $U_d(r_b) = \Gamma\pi(r_b^2 - r_i^2)$ invested in delaminating the gel-glass interface with energy density $\Gamma$, starting from an initial basal radius of the biofilm $r_i$ to its final basal radius $r_b$, and (2) the elastic energy stored in the gel $U_e(r_b, V) = \mu r_b^3 f(V/r_b^3)$, where $\mu$ is the shear modulus and $f = f(V/r_b^3)$ is the dimensionless elastic potential energy as a function of dimensionless volume, obtained from finite element simulations. Frictional forces come into play only after the gel begins to delaminate and the biofilm expands on the substrate, which we model using the *Rayleigh dissipation function,*

$D(r_b, \dot{r}_b) = \frac{1}{2}\int_0^{r_b}\int_0^{2\pi} \eta |\boldsymbol{v}(r; r_b, \dot{r}_b)|^2 r\, dr d\theta$, where $\eta$ is the friction coefficient. The Euler-Lagrange equation for this system, with the generalized coordinate $r_b \equiv r_b(V)$ is written as $\frac{\partial U}{\partial r_b} = -\frac{\partial D}{\partial \dot{r}_b}$ which gives

$$\text{H}\frac{d\tilde{r}_b}{d\tilde{V}} = \text{M}\frac{\tilde{F}(\tilde{V}/\tilde{r}_b^{\,3})}{\tilde{V}} - \frac{1}{\tilde{V}\tilde{r}_b}, \tag{1}$$

where $\tilde{r}_b \equiv r_b/r_i$ and $\tilde{V} \equiv V/r_i^3$ are dimensionless quantities and $\tilde{F}(x) = xf'(x) - f(x)$ (Supplementary Information). Crucially, the biofilm growth dynamics are governed by two dimensionless variables that emerge naturally from this formulation: the dimensionless friction $\text{H} = \eta g r_i^2/4\Gamma$ and the dimensionless elastic modulus $\text{M} = 3\mu r_i/2\pi\Gamma$ (where $g$ is the biofilm growth rate), which measure the relative importance of frictional dissipation and elastic potential energy to interfacial energy, respectively. Solving (1) we find that as the volume increases, the system initially exhibits cavitation-like expansion where biofilm growth induces only elastic deformation in the gel with no sliding motion of the biofilm with respect to the substrate, along with a growing contact angle[31]. The system then transitions to delamination where breakage of interfacial bonds between the gel and the glass substrate becomes energetically favorable and leads to sliding of the biofilm cells along the substrate. In this limit, biofilm growth mimics a "hydraulic" fracture, which gives rise to a decreasing contact angle and lens-shaped biofilms. Finally, as the biofilm continues to grow, the system transitions to a *friction-limited delamination* regime where friction retards expansion on the substrate leading to a growing contact angle again and hence dome-shaped biofilms (Fig. 2d). Experimentally, the observed contact angles of biofilms are controlled by $V$, $\eta$ and $\mu$. Benchmarked by experimentally measured values (Supplementary Information), the theoretical phase diagram closely matches those attained experimentally and recapitulates many salient features (Fig. 2c, Extended Data Fig. 5 and 6). In the small $\eta$ limit, the model reduces to the previous interfacial cavitation model[31] where the shape is independent of $\eta$. In the large $\eta$ limit, this model predicts that the shape transition occurs at a constant ratio of $\eta$ to $\mu$, consistent with the agent-based simulations quantitatively and with the *bap1*-titration experiment qualitatively; in this limit, the energetics is dominated by the balance between the frictional dissipation and elastic deformation of the gel, so the shape only depends on $H/M = (\pi g r_i/6)(\eta/\mu)$.

**Boundary evolution determines cell trajectories and positional fate**

As a biofilm grows, the dwelling cells can self-organize spatially and temporally; we therefore considered the implications of the different morphologies on the internal structural evolution of the biofilms. Critical to understanding this self-organization process is revealing the trajectories of different cells inside the biofilm. To this end, we used a bacterial strain in which each cell contained a single bright punctum[32,33] that we tracked over time (Fig. 3a). We projected all trajectories into the axisymmetric coordinates of the biofilm (Fig. 3b, grey lines) and overlayed them with the averaged trajectories for cells that ended near the boundary (Fig. 3b, purple lines). This allowed us to visualize the spatial distribution of different cell lineages.

Under soft confinement, cell trajectories followed a fountain-like flow pattern where cells that originated near the core left the substrate and overtook frictionally slowed cells near the substrate (Fig. 3b, d), in a manner similar to unconfined biofilms[32]. In contrast, when the gel was stiff, all cell trajectories bended *upward* away from the substrate (Fig. 3b); as a direct consequence, the basal layer of the biofilm consisted mainly of cell lineages that always stayed on the substrate (Fig. 3d). A similar change in cell trajectories was observed in the ABSs upon changing biofilm morphology, therefore ruling out biological signaling as the cause of the observed change in cell trajectories (Extended Data Fig. 7).

We hypothesized that the observed alteration in cell trajectories was driven by the differing progressions of the biofilm-gel boundary. This is because the cells at the boundary are anchored to the gel[20], requiring the cells to track the motion of that material point. To support this hypothesis, we tracked the displacements of the boundary by embedding and tracking tracer particles in the agarose gel. Consistent with our theoretical model for the overall shape, we found two distinct modes of tracer trajectories corresponding to the dome-shaped and lens-shaped modes of growth (Extended Data Fig. 8). In the stiff gel (lens-shaped) regime, the tracers were displaced vertically away from the substrate as the gel delaminated to continuously create new biofilm-gel and biofilm-glass interfaces; in contrast, in the soft gel (dome-shaped) regime, little new biofilm-glass interface was created and instead the biofilm-gel boundary expanded to accommodate cell proliferation. To reveal the creation of new interfaces in the stiff gel regime, we mapped the "age" of the biofilm-gel interface (Fig. 3c) and found that the central part of the interface was indeed older as it was created earlier on during biofilm growth. Because a cell adhered to the gel boundary will track the boundary displacement, this naturally leads to upward bending of the cell trajectories in the lens-shaped limit (Fig. 3e). Lending support to this argument, when we deleted the key exopolysaccharide biogenesis gene *vpsL* such that cells were not adhesive to the boundary[20], the cell trajectories no longer bent upwards despite the fact that the biofilm was similarly lens shaped (Extended Data Fig. 9). Therefore, we conclude that progression of the biofilm-gel boundary combined with the cell-gel adhesion determines the positional cell fate and the spatiotemporal distribution of lineages within a biofilm.

**Morphology-induced nematic structural transition**

A hallmark of LCs is the self-organization of orientational order due to anisotropic interparticle interactions. The ground state of an unconfined nematic assumes a constant scalar order parameter $S(r)$ and a uniform director $\hat{n}(r)$[34]; when a nematic is confined, however, the anchoring condition at the boundary can often lead to geometric frustrations and creation of topological defects[35,36]. Given the elongated shape of *V. cholerae* cells, a natural question is to understand how the evolving biofilm boundary influences the orientational order inside the biofilm. To quantify the orientational order, we measured the spatially varying nematic order parameter tensor $Q(r) = \langle 3\hat{n}_i \otimes \hat{n}_i - I \rangle/2$ where angled brackets denote spatial averaging across cell orientations $\hat{n}_i$ of different cells $i$ in a local neighborhood (Fig. 4 and 5). The scalar order parameter $S$ was defined as the maximum eigenvalue of $Q$, and the nonpolar director $\hat{n}(r) = -\hat{n}(r)$, which marked the averaged local orientation of cells, was the corresponding eigenvector. Under this definition, when $S = 1$ cells are perfectly aligned parallel to each other and when $S = 0$ the cells are isotropically disordered.

We measured $S(r)$ and $\hat{n}(r)$ in 6-18 biofilms at each gel concentration and averaged them to generate "prototypes" of biofilm organization and obtain the overall nematic order $\bar{S}(r)$ and $\bar{n}(r)$, as shown in Fig. 4 and 5. As the gel concentration increased from 0.2% to 1.5%, $\bar{S}(r)$ of the dome-shaped biofilm increased. In general, $\bar{S}(r)$ was maximized at the origin, where $\bar{S} \approx 0.6$ and gradually reduced to as low as 0.4 as $r$ increased but increased again to ~0.6 at the boundary (Fig. 4). For the higher gel concentrations ($c = 1.0\%$ and 1.5%), the director field followed a "bipolar" structure with two surface defects, called boojums, sitting at the origin and apex of the biofilm, similar to those observed in thermotropic LCs confined in a spherical droplet with planar anchoring

(Fig. 5)[35]. Concomitant with the dome-to-lens shape transition at $c = 1.5\%$, we observed a marked topological transition of the director field. In lens-shaped biofilms, the two boojums remained and the director connecting the two boojums bent smoothly in the middle of the biofilm, allowing us to identify a bipolar ellipsoid (Fig. 5b). However, an additional -1/2 *disclination loop* emerged around this ellipsoid, making it topologically distinct from the dome-shaped biofilms. $\bar{S}$ was generally higher in lens-shaped biofilms but still showed a small dip in the interior (Fig. 5c).

We attribute the topological difference of the director field in the biofilms to the distinct behavior at the biofilm-gel-glass triple contact point, which, in turn, is controlled by the overall biofilm morphology explained above. While the dome-shaped biofilms support a bipolar structure, maintaining the same topological organization in lens-shaped biofilms would require significant bending deformation of the director field as shown in the hypothetical configuration in Fig. 5d. To alleviate the large distortion in the director field, the cells instead adopt a splay configuration to fill the wedge-shaped triple contact point in lens-shaped biofilms, which topologically necessitates the emergence of a -1/2 disclination loop (leading to a dip in $\bar{S}$ in the interior of the biofilm). This is analogous to the formation and dissociation of pairs of $\pm 1/2$ disinclinations which occur in many other active nematic systems, such as in microtubule suspensions[1] and living liquid crystals[37]. In the growing lens-shaped biofilms, the +1/2 disclination is anchored at the triple contact line (Fig. 5d)[38].

An important contrast between the growing nematic system studied here and equilibrium nematics is the correlation of $S$ with the deformation in the director field. In equilibrium nematics, large director deformation energy at the topological defect core competes with the anisotropic interactions between LC molecules and reduces $S$ in the nearby region[39,40]. In contrast, in biofilms stronger ordering was found at the core of the boojum-like defects compared to the bulk. We also found that mutants lacking cell-substrate adhesins grew into macroscopically disordered, dome-shaped biofilms, highlighting the importance of cell-boundary interactions, in addition to anisotropic cell-cell potentials[15], in driving macroscopic ordering (Extended Data Fig. 10). Therefore, we suggest that the observed macroscopic organization arises from mechanical interactions between biofilm cells and the boundaries, which then propagate into the interior of the biofilm. Collective alignment occurs along the basal layer, where growth-induced stress drives cell verticalization at the core and radial alignment in the rim, creating the boojum near the origin[14]. In parallel, as the biofilm grows and the biofilm-gel interface area dilates, it exerts a radial extensile stress to the gel, which, in turn, aligns the boundary-adhered cells radially leading to the emergence of another boojum at the apex of the biofilm[20]. Cell ordering at the top and bottom boundaries propagates into the interior of the biofilm, defining the internal organization of the cells.

Consistent with the picture of boundary and stress-driven alignment, we observed that: first, the distance the nematic order propagates into the interior increased with increasing gel stiffness and therefore stress, leading to overall higher $\bar{S}$. Second, when we removed the boojum defect at the biofilm-substrate interface by deleting the adhesins, the -1/2 disclination loop disappeared and the boundary-driven alignment from the top penetrated deeper into the biofilm (Extended Data Fig. 10). Finally, when we imposed degenerate planar anchoring on the boundaries (by using the Δ*vpsL* mutant where cells aligned parallel to the boundary), the biofilm consisted of purely horizontally aligned cells in the interior, similar to classical bacterial colonies[9,41]. To sum up, our results demonstrate how biofilm shape, set by macroscopic energetics, drives the emergence of distinct long-range nematic ordering in the interior of biofilms through mechanical coupling with the boundaries.

**Conclusion**

Understanding the different modes of biofilm growth is critical to understanding how biofilms and, more generally, how growing organisms can alter their morphology and internal architecture in response to environmental signals and constraints. This is also useful when engineering new classes of growing active materials that adapt to their surroundings by considering changes in geometry and internal organization. Here we showed how a growing biofilm actively shapes its environment and its internal architecture and lineages through mechanical coupling to its surroundings. Because the geometry and directionality of cell growth dictates the accessibility of nutrients to the entire biofilm, nematic cell ordering may afford improved nutrient/waste diffusion into and out of the innermost portions of the biofilm[42]. Different cell trajectory patterns will transport different lineages to separate regions of the biofilm and determine where antibiotic tolerant or persistent cells end up spatially in a biofilm[43]. The variation in cell positional fate could further couple with heterogeneous gene expression pattern in a biofilm leading to segregation of cells with different internal states. From an application point of view, the phenomena discovered here could offer new ways to mechanically guide biofilm growth, leading to new strategies to suppress the growth of harmful biofilms, and to design and program beneficial ones. Finally, while in this work we manipulated the extracellular matrix production through mutagenesis, it is intriguing to consider how and when bacteria adapt extracellular matrix production to mechanical cues through gene regulation to guide their own development.

**Acknowledgments**: Research reported in this publication was supported by the National Institute of General Medical Sciences of the National Institutes of Health under Award Number DP2GM146253. J.Y. holds a Career Award at the Scientific Interface from the Burroughs Wellcome Fund (1015763.02). This publication was made possible in part with the support of the Charles H. Revson Foundation (J.N). J.-S.B.T. is a Damon Runyon Fellow supported by the Damon Runyon Cancer Research Foundation (Grant No. DRG-2446-21). T.C. acknowledges the support of Dr. Timothy B. Bentley, Office of Naval Research Program Manager, under award number N00014-20-1-2561.

**Author contributions:** J.N. and J.Y. conceptualized the project. J.N. and Q.Z. performed the experiments and J.N. and J.-S.B.T. performed the data analysis. J.N., M.K., T.H., S.Z., T.C. and J.Y. formulated the theoretical model. C.L. and S.Z. developed the agent-based simulations. All authors contributed to the writing of the manuscript.

**Competing interests:** The authors declare that they have no competing interests.

**Supplementary Materials:** Supplementary information is available for this paper.

Correspondence and requests for materials should be addressed to either [jing.yan@yale.edu](mailto:jing.yan@yale.edu) or [talco@mit.edu](mailto:talco@mit.edu)

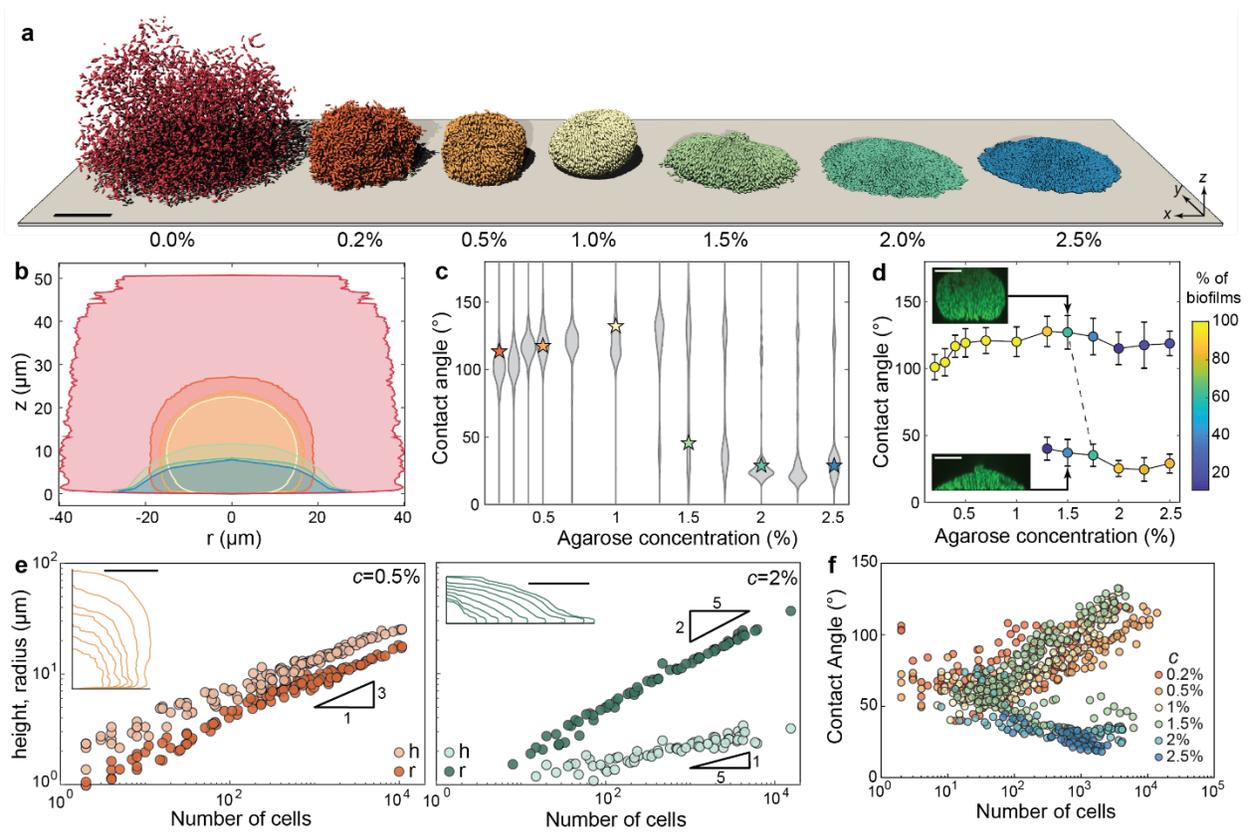

**Figure 1: Biofilm shape bifurcation in response to environmental stiffness| (a)** Reconstructed biofilms grown under agarose gels with different concentrations. Biofilms consist of $8600 \pm 700$ (mean $\pm$ s.d.; range $7245 - 9420$) cells. **(b)** Shape of biofilms in (a) in cylindrical coordinates. The contours are reflected about $r = 0$. **(c)** Violin plot of contact angles calculated for biofilms grown under different agarose concentrations. Each chord represents a probability distribution function calculated from $136 \pm 53$ (mean $\pm$ s.d.; range $58 - 269$) mature biofilms. Stars correspond to biofilms shown in (a) and (b). **(d)** Bifurcation of the biofilm contact angle with agarose concentration. Each point (and error bar) corresponds to the peak (and standard deviation) of a local gaussian fit that encompasses all biofilms with contact angles either greater than or less than 75°. Inset: two examples of mature biofilms with different morphologies grown under 1.5% agarose gels. **(e)** Plot of the maximum height and maximum radius of biofilms grown under a 0.5% gel (*left*) and 2% gel (*right*). Data corresponds to the ensemble of 12 and 6 different biofilms

imaged over time, respectively. Inset: shape evolution of a single biofilm under each condition. (**f**) Time-evolution of the contact angle for biofilms grown under gels with different stiffnesses. Scale bars, 10 μm.

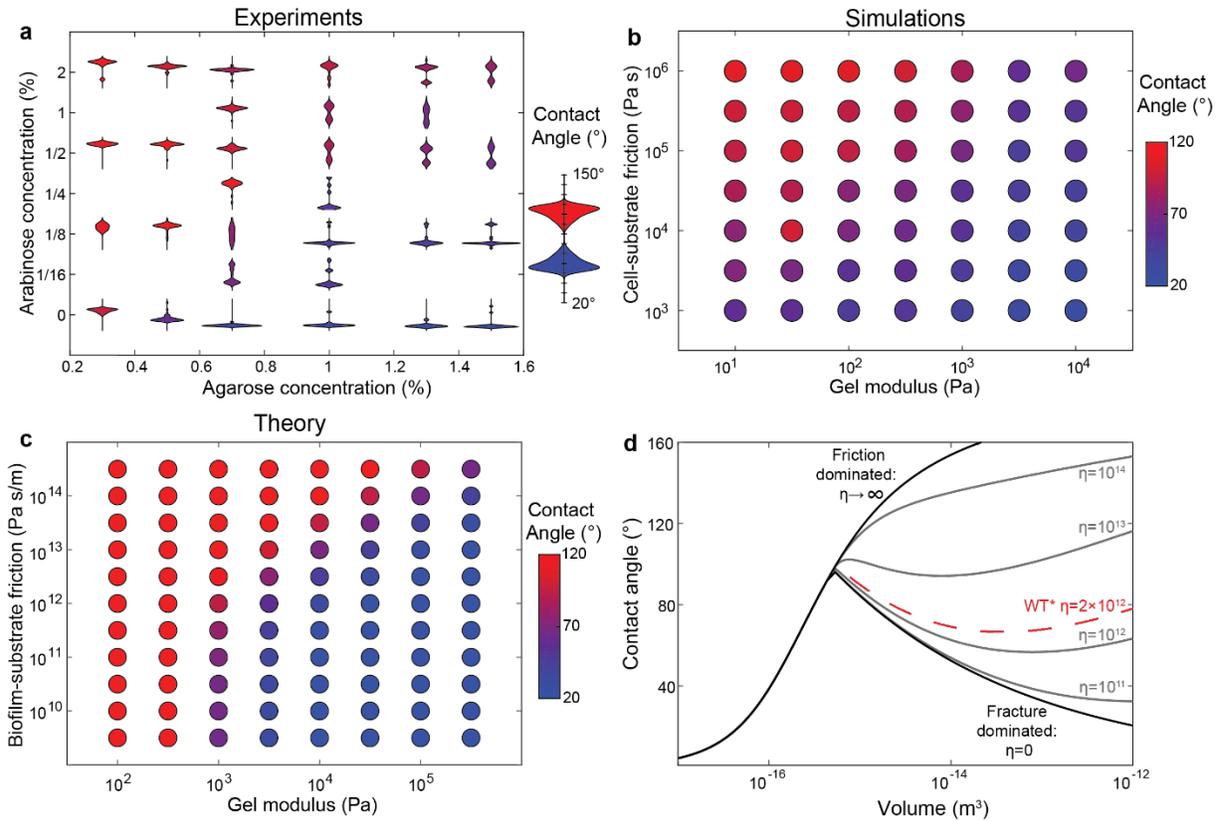

**Figure 2: Environmental stiffness and biofilm-surface adhesion jointly control biofilm shape|** **(a)** Phase diagram showing the experimental distribution of biofilm shapes for cells producing varying amounts of the surface adhesion protein *bap1*, controlled by an arabinose inducible promotor, and grown under different stiffness environments. Each icon corresponds to a violin plot of contact angles with red to blue corresponding to large and small mean contact angles, respectively. Each histogram corresponds to $41 \pm 25$ biofilms (mean $\pm$ s.d.; range $6 - 138$). **(b)** Phase diagram showing biofilm contact angles from agent-based simulations for different cell-substrate friction coefficients and gel stiffnesses. Each dot corresponds to a single unique simulation. **(c)** Phase diagram showing predicted biofilm contact angles calculated from the continuum model (Supplementary Information) for $V = 10^{-13}$ m$^3$. **(d)** Predicted evolution of the contact angle with growing volume for stiffness $\mu = 3$ kPa for different friction coefficient $\eta$.

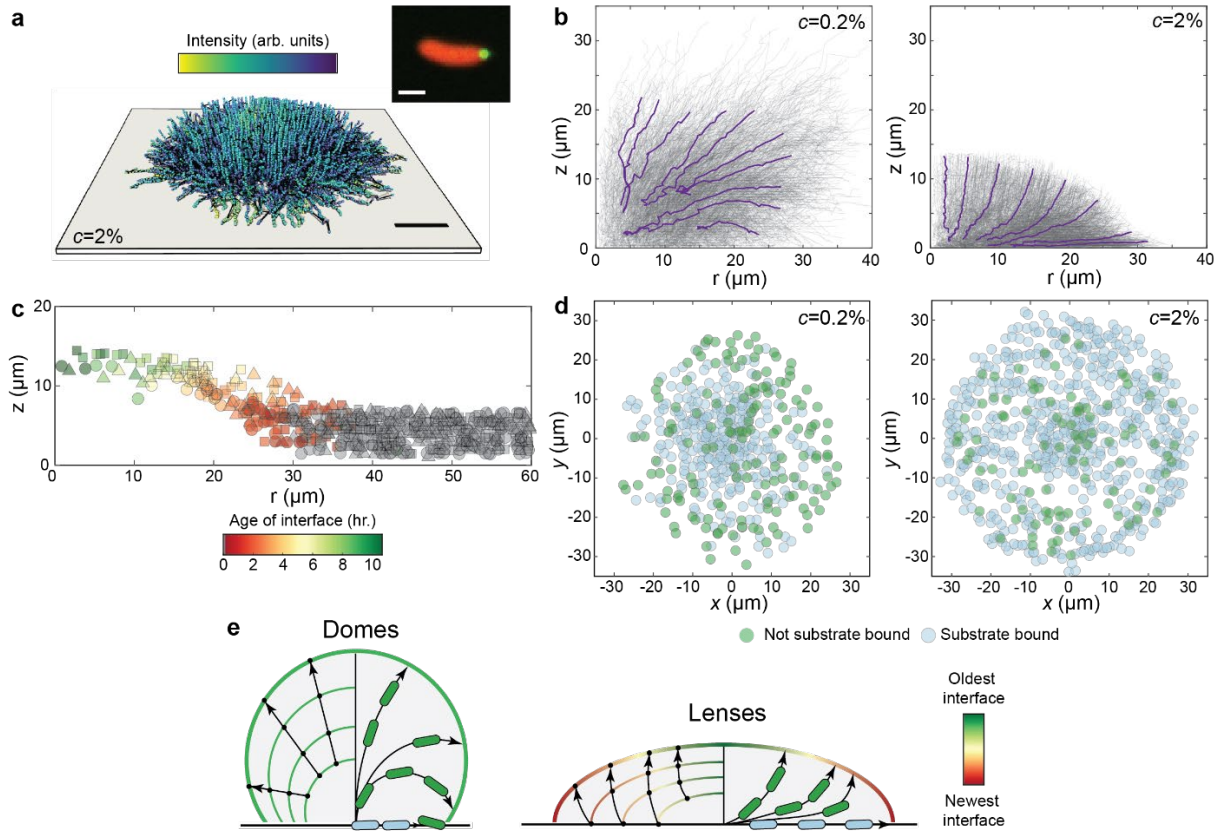

**Figure 3: Boundary conditions dictate cell fate in biofilm|** (**a**) Reconstructed cell trajectories from puncta tracking in a biofilm confined by a stiff gel ($c = 2\%$). Colors denote the intensity of the fluorescently labelled puncta. Scale bar, 10 μm. Inset: image of a green-punctum-containing red *V. cholerae* cell. Scale bar, 1 μm. (**b**) Puncta trajectories from biofilms grown under two different conditions projected into $(r, z)$ space. Purple lines denote averaged trajectories that end near the edge of the biofilm. (**c**) Age of the biofilm-gel interface measured by tracking the displacement of tracer particles embedded in the agarose gel. The delamination time, i.e. birth of the local interface, is defined as the time point when the vertical displacement of the corresponding tracer particle exceeds 0.5 μm. Data consists of an ensemble of three different biofilms labelled with three different markers. (**d**) Basal layer puncta labelled by whether its height has exceeded 3 μm or not during its entire history, corresponding to cells that have transiently left the surface

(green) and cells that are always substrate bound (blue), respectively. (**e**) Schematic representation of the cell trajectories and their coupling to boundary evolution.

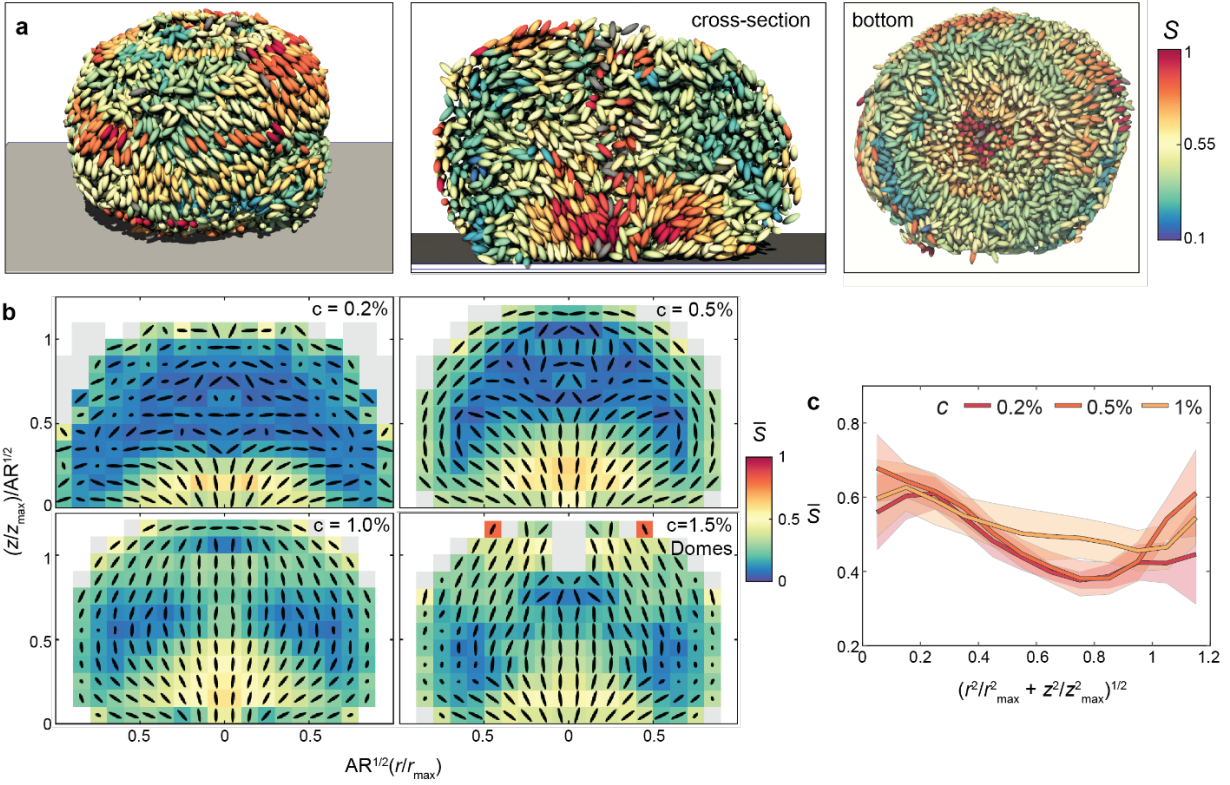

**Figure 4: 3D spatial variation in cell orientations and ordering in dome-shaped biofilms** (a) Three-dimensional reconstruction of a biofilm grown under soft confinement ($c = 0.5\%$). Cells are colored based on the scalar order parameter calculated in each differential volume with $\Delta r = 2$ μm, $\Delta z = 2$ μm, $\Delta \theta = 45°$. (b) Azimuthally averaged cell orientations for biofilms grown in different stiffness environments. Colors denote the scalar order parameter and the ovals denote the average director of the cells projected into $(r, z)$ space. Data is first averaged azimuthally in each biofilm then averaged across $13 \pm 5$ (mean $\pm$ s.d.; range 5-18) different biofilms. To account for different sizes of biofilms, $r$ and $z$ were rescaled by $r_{max}$ and $z_{max}$ prior to averaging and rescaled after averaging such that the aspect ratio $AR$ was equal to the mean aspect ratios of the underlying biofilms. Note the data shown are reflected about $r = 0$. Grey denotes regions with an insufficient number of cells for averaging. (c) Scalar order parameter averaged as a function of the normalized distance to the origin (mean $\pm$ s.d.). For each condition, data is first averaged in each biofilm and then averaged across biofilms (data corresponds to the same underlying data as b).

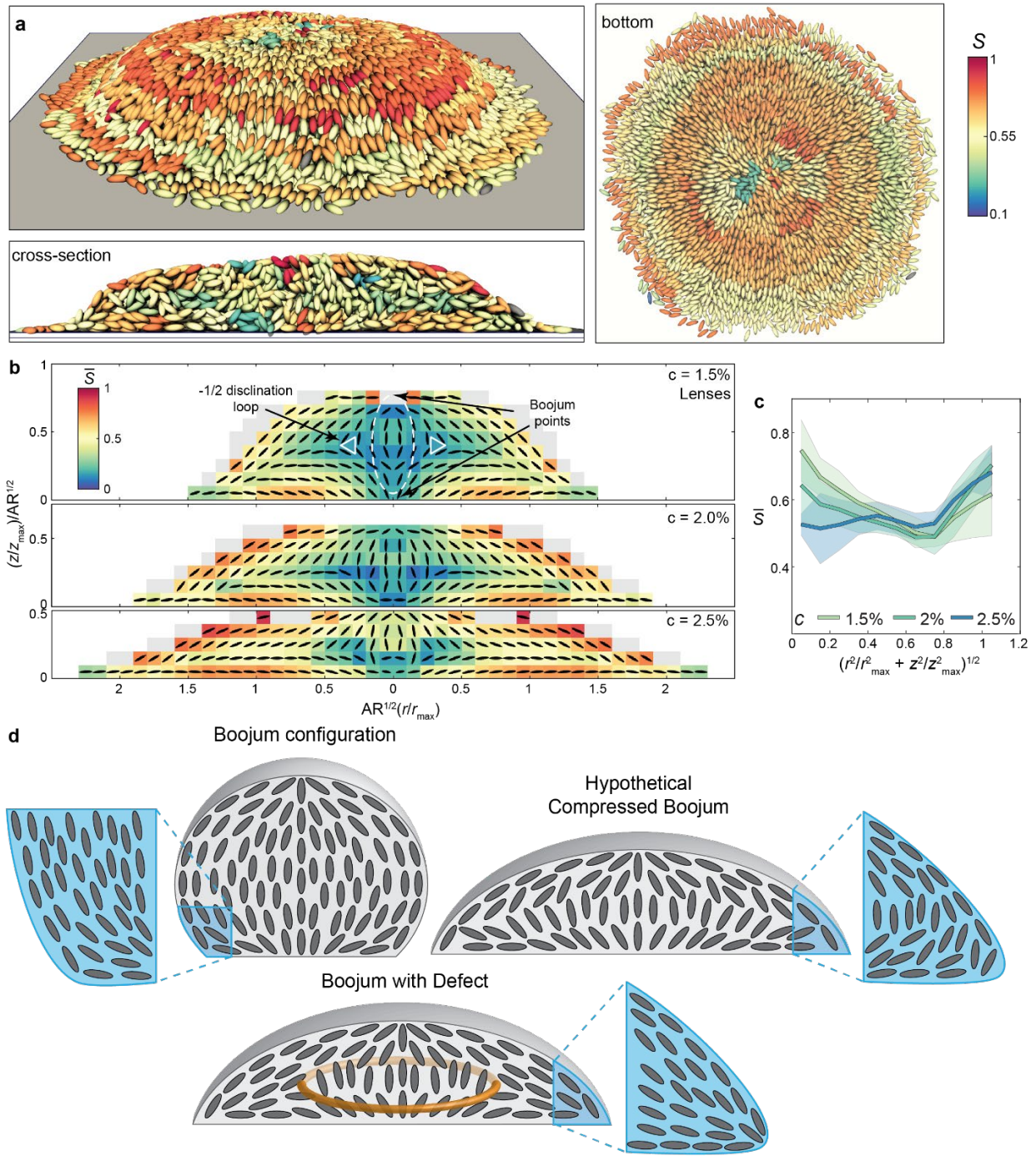

**Figure 5: 3D spatial variation in cell orientations and ordering in lens-shaped biofilms** (**a**) Three-dimensional reconstruction of a biofilm grown under stiff confinement ($c = 2\%$). Cells are colored based on the scalar order parameter calculated in each differential volume with $\Delta r = 2$ μm, $\Delta z = 2$ μm, $\Delta \theta = 45°$. (**b**) Azimuthally averaged cell orientations for biofilms grown in different stiffness environments. Colors denote the scalar order parameter, and the ovals denote the average director of the cells projected into $(r, z)$ space. Data is first averaged azimuthally in each biofilm then averaged across $11 \pm 4$ (mean $\pm$ s.d.; range 6-16) different biofilms. Grey denotes regions with an insufficient number of cells for averaging. (**c**) Scalar order parameter averaged as a function of the normalized distance to the origin (mean $\pm$ s.d.). (**d**) Schematic of how shape controls nematic organization inside a biofilm. A hypothetical compressed Boojum configuration will have significant distortion in the local director field near the triple contact point, which is alleviated by a continuous splay conformation at the edge and the formation of a $-1/2$ disclination loop in the interior, as we see in biofilm grown under stiff confinement.

Supplementary Information for

# Biofilms as self-shaping growing nematics

Japinder Nijjer[1], Mrityunjay Kothari[2], Changhao Li[3], Thomas Henzel[2], Qiuting Zhang[1], Jung-Shen B. Tai[1], Shuang Zhou[4], Sulin Zhang[3,5], Tal Cohen[2,6], Jing Yan[1,7]

**Methods**

Growth and imaging of confined biofilms

The strains used in these experiments were derivatives of the C6706 El Tor strain and contained a point mutation in the diguanylate cyclase *vpvC* (*vpvC*$^{W240R}$), which caused upregulated c-di-GMP production and therefore constitutive biofilm production[1]. Unless otherwise noted, these strains also included a deletion of the *rbmA* gene to isolate the effects of cell-cell adhesion[2,3] – this strain background was labelled WT*. In addition to the WT* strain, we also worked with a set of strains in which the genes encoding the adhesins Bap1 and RbmC were deleted. The Δ*bap1*Δ*rbmC* mutant exhibited little friction as it grew on the substrate[4]. We also utilized a set of strains that did not produce any extracellular polysaccharides through the deletion of the Vibrio polysaccharide synthase gene *vpsL*[5], which behaved analogously to other non-biofilm-forming bacteria[6–9]. The bacteria were also genetically modified to constitutively express the fluorescent protein mNeonGreen, or in the case of cell trajectory measurements, mScarletI. For cell trajectory measurements, we used a strain containing mNeonGreen fused to the µNS protein from the avian reovirus which self-assembled to form a single intracellular punctum. These puncta are inherited by one of the single daughter cells while a new one self-assembles in its sibling, thereby allowing the tracking of individual lineages over time[10]. For a complete list of strains see Table S1.

Biofilm growth experiments were performed in M9 minimal media (Sigma Aldrich) supplemented with 0.5% glucose (Sigma Aldrich), 2 mM MgSO4 (JT Baker) and 100 µM CaCl2 (JT Baker) (henceforth referred to as M9 media). For confined growth experiments, cells were first grown under shaken conditions overnight in LB broth (BD). The overnight culture was back-diluted 30× in M9 media and grown under shaken conditions until an optical density (OD) of 0.05-0.25 (approximately 2 hours). Concurrently, agarose polymer (Invitrogen) of a given concentration was boiled in M9 media and then placed in a water bath to cool to 40-50°C without gelation. The bacterial culture was diluted in M9 media to an OD of 0.001-0.003 and a 1 µL droplet of this diluted culture was deposited in the center of a glass-bottomed 96 well plate (MatTek). The droplet was covered with 20 µL of the liquid agarose, which quickly solidified at room temperature and sandwiched the bacteria between the solidified gel and the glass substrate (note we neglect the ~5% dilution of the agarose by the droplet). Finally, 200 µL of M9 media was added in the well on top of the solidified agarose to act as a nutrient reservoir. Cells were finally grown under static conditions at 30°C and imaged at various times during development.

Overview of imaging and image analysis

Imaging was performed using a Yokogawa CSU-W1 spinning-disk confocal scanning unit mounted on a Nikon Ti2-E microscope body, using the Nikon perfect focus system and images were acquired using Nikon Elements 5.20. For high-resolution, single-cell level imaging, a 100× silicon oil immersion objective (Lambda S 100XC Sil, numerical aperture = 1.35) was used. At low agarose concentrations ($c \leq 0.5\%$) a z-step size of 0.195 µm was used, while at high agarose concentrations ($c > 0.5\%$) a z-step of 0.13 µm was used. For high-throughput biofilm morphology measurements, a 60× water immersion objective (CFI Plan Apo 60XC, numerical aperture = 1.20) and a z-step size of 0.4 µm was used. The green mNeonGreen fluorophore was excited using a 488 nm laser, the red mScarletI fluorophore was excited using a 561 nm laser and the far-red fluorescent beads were excited using a 640 nm laser. For time course imaging, cells were incubated in a Tokai-Hit stage-top incubator at a temperature of 30°C.

After acquisition, images were deconvolved using Huygens 20.04 (SVI). The high-resolution single-cell images were then segmented into individual cells using methods described elsewhere[4,11]. Briefly, the images were first binarized layer-by-layer using an adaptive Otsu method and the cells were then segmented using an adaptive thresholding scheme. The cell locations and directors where then determined from the center of mass and the principal axis from a principal component analysis of the segmented voxels, respectively. We further defined a cylindrical coordinate system where the origin was set by finding the radial center of mass of all the segmented points.

High-throughput contact angle measurements

To attain high-throughput measurements of the contact angles across many biofilms, a large, tiled image approximately 1 mm × 1 mm, containing $27 \pm 20$ (range 4-87) biofilms was first taken at 2-6 hours after seeding to identify bacteria which started at the gel-substrate and then taken at 12-20 hours after seeding for contact angle ($\psi$) measurement. Since we were only interested in measuring the effective $\psi$, we restricted our attention to the bottom 5 µm of each biofilm in this assay. The images were deconvolved and then segmented using a custom Matlab (2018a) script. First the images were denoised and binarized layer-by-layer using a Wiener 2-D adaptive noise-removal filtering and Otsu thresholding. Biofilms were then either automatically or manually identified as large, connected binarized voxels. For each biofilm and for each layer, a convex hull that contained all binarized pixels was found and the area of the hull was taken to be the cross-sectional area of the biofilm $A(z)$ at each height $z$. From the cross sectional area the effective radius was calculated as $r(z) = (A(z)/\pi)^{1/2}$. The contact angle was then found by fitting a linear slope and calculating $\psi = \tan^{-1}(dr/dz) + 90°$ (Extended Data Fig. 2).

Tracing of cell trajectories

To trace the cell lineage trajectories in the biofilm, we tracked the trajectories of individual puncta inside the biofilm. First, the deconvolved images were registered using MATLAB built-in functions to minimize frame-to-frame jitter. Individual puncta were then identified as local maxima in the images, and subpixel resolution was attained by fitting a parabola around the maxima. This process was repeated for all frames and the particles were connected over time using TrackMate particle tracking software[12]. These puncta trajectories were projected into the cylindrical coordinates of the biofilm. The averaged trajectories were calculated by averaging all trajectories whose final $(r, z)$ was within 3 μm of the target final coordinate $(r_{tar}, z_{tar})$. We neglected parts of the averaged trajectories where fewer than 3 trajectories were averaged. The final target coordinates were chosen at different points near the boundary.

Visualization of the gel deformation

To visualize the deformation of the agarose gel, we diluted 200 nm far-red fluorescent particles (Invitrogen) at a ratio of 1/100 into the molten agarose gel prior to encasing the bacteria. In the first step of the data analysis process, a portion of the deconvolved images where little particle motion was expected was used to register the images using MATLAB built-in functions. Using a procedure similar to puncta tracking, the fluorescent particles were identified and tracked by finding local maxima and using the TrackMate particle tracking software. The "age of the interface" was determined by finding the time when the vertical displacement of particles near the substrate (initially within 5 μm) exceeded a threshold value of 0.5 μm, corresponding to a local delamination event.

Quantification of cell ordering

To quantify the average cell ordering inside the biofilms we averaged cell directors using the $\mathbf{Q}$-tensor model of liquid crystals[13]. For each cell $i$, we first converted the director into a head-tail symmetric quantity by taking the inner product of the director with itself $\mathbf{Q}_i = (3\hat{\mathbf{n}}_i \otimes \hat{\mathbf{n}}_i - \mathbf{I})/2$ (where $\hat{\mathbf{n}}_i$ is in cartesian coordinates). Each biofilm was discretized into cylindrical sectors with $\Delta r = 2$ μm, $\Delta z = 2$ μm and $\Delta \theta = \pi/4$ and $\mathbf{Q}$ was then averaged in each sector yielding a locally averaged, spatially varying nematic order parameter $\mathbf{Q}(r, \theta, z)$. To azimuthally average $\mathbf{Q}$, we first converted it to cylindrical coordinates through the transformation $\mathbf{Q}_p = \mathbf{R}^T \mathbf{Q} \mathbf{R}$, where $\mathbf{R}$ is the transformation matrix, and then averaged across $\theta$. Finally, to average across many biofilms at the same agarose concentration, we rescaled each biofilm by its maximum radius and height, yielding $\mathbf{Q}_p \left( \frac{r}{r_{max}}, \frac{z}{z_{max}} \right)$ and then averaged $\mathbf{Q}_p$ across many biofilms. To visualize and quantify the nematic order parameter, we calculated the scalar order parameter $S$ as the maximum eigenvalue of $\mathbf{Q}$ or $\mathbf{Q}_p$ and $\hat{\mathbf{n}}$ as the corresponding eigenvector.

Agent-based simulations

The agent-based simulations are built upon those developed in Nijjer et al.[4] and updated to include cell-gel adhesion. Briefly, we modelled the bacteria as spherocylinders that repelled each other through Hertzian mechanics and exhibited JKR-like adhesion to the glass substrate and confining gel. As cells moved relative to the substrate, they experienced a frictional force proportional to $\eta_{ABS} v$ where $v$ was the velocity of the cells (driven by growth-induced expansion). The gel was modelled using a particle representation such that it had linear elastic properties with stiffness $\mu_{ABS}$ as well as adhesion to the glass substrate. The simulations began with a single cell which grew with a growth rate $g$ and divided upon doubling in length. The simulation was stopped after 10-13 generations. The full details of the simulation will be reported elsewhere.

Continuum modelling of biofilm shape morphogenesis

Here we present a minimal model to explain the macroscopic morphogenesis of *V. cholerae* biofilms confined between an infinite elastic material bonded to a soft substrate. As the biofilm grows, it induces elastic deformation in the surrounding gel, which can lead to delamination of the gel from the substrate. We assume the biofilm is axisymmetric and a schematic of the geometry is given in Extended Data Figure 5.

We take the stiffness of the gel to be $\mu$ and assume an initial defect of radius $r_i$ where the gel is initially not adhered to the substrate. We treat the growth of the biofilm as a quasistatic process and assume the gel is always in mechanical equilibrium. For a given biofilm volume $V$ the biofilm has shape $r(z)$ with basal radius $r_b \geq r_i$; the elastic energy due to deformation of the gel is $U_e = \mu r_b^3 f\left(\frac{V}{r_b^3}\right)$, where $f\left(\frac{V}{r_b^3}\right)$ quantifies the elastic potential energy, which is only a function of the dimensionless volume $V/r_b^3$. There is no simple analytical form for $f$ so we instead numerically determine it using the Finite Element software ABAQUS/CAE 2017[14]. The energy release due to breakage of gel-substrate bonds is $U_d = \Gamma\pi(r_b^2 - r_i^2)$ with energy density $\Gamma$, for $r_b > r_i$ and 0 otherwise. Taking these two contributions together, the total energy of the system for $r_b \geq r_i$ is

$$U = U_d + U_e = \Gamma\pi(r_b^2 - r_i^2) + \mu r_b^3 f\left(\frac{V}{r_b^3}\right). \tag{1}$$

As the biofilm grows, it also experiences friction as cells are advected along the substrate due to growth-induced motion. This is because the secreted adhesins bond the cells to the substrate leading to frictional behavior as relative motion occurs[15]. We assume the frictional force is proportional to the velocity such that the friction experienced due to motion of a basal point on the biofilm is $\boldsymbol{F}_f = \eta\boldsymbol{v}$, where $\boldsymbol{v}$ is the velocity of the biofilm at the substrate and $\eta$ is the friction coefficient. The basal velocity is nearly axisymmetric and has little azimuthal component. Therefore, integrating over the whole basal area of the biofilm, the corresponding Rayleigh dissipation function is

$$D = \pi \int_0^{r_b} \eta |\boldsymbol{v}|^2 r \, dr. \tag{2}$$

In this case, we write $\boldsymbol{v} = v_r(r)\hat{r}$, and the velocity satisfies the symmetry condition $v_r(0) = 0$ and the kinematic condition $v_r(r_b) = dr_b/dt$. We neglect energy dissipation in the bulk of the biofilm as the dissipation from growth along the substrate dominates[16,17]. Because the basal layer of the biofilm has a uniform density[4], mass conservation requires that the velocity satisfies

$$\frac{1}{r}\frac{d(rv_r)}{dr} = g', \tag{3}$$

where $g'$ is the effective basal growth rate, which accounts for the loss and accumulation of biomass from the neighboring cells above and is not necessarily equal to the intrinsic biofilm growth rate $g$. Assuming $g'$ is spatially uniform, and solving equation (3), we find

$$v_r = \frac{dr_b}{dt}\frac{r}{r_b}. \qquad (4)$$

Note that in reality the velocity profile in the basal layer of WT* biofilms may deviate from this simple linear form;[4] however, the exact functional form likely does not affect the main conclusions of the current analysis. Substituting into (2), the Rayleigh dissipation function becomes

$$D = \frac{\pi \eta r_b^2 \dot{r}_b^2}{4} \qquad (5)$$

where $\dot{r}_b = dr_b/dt$. Writing the Euler-Lagrange equation $\frac{\partial D}{\partial \dot{r}_b} = -\frac{\partial U}{\partial r_b}$ for the generalized basal radius coordinate $r_b$ yields

$$\frac{\eta \pi r_b^2}{2}\frac{dr_b}{dt} = 3\mu r_b^2 \left(\frac{V}{r_b^3} f'(V/r_b^3) - f(V/r_b^3)\right) - 2\Gamma \pi r_b. \qquad (6)$$

Using the fact that the whole biofilm grows exponentially ($dV/dt = gV$), we rewrite $\frac{dr_b}{dt} = \frac{dr}{dV}gV$. Substituting into (6) yields an ordinary differential equation for the evolution of the basal radius,

$$\frac{\eta \pi r_b^2 gV}{2}\frac{dr_b}{dV} = 3\mu r_b^2 \left(\frac{V}{r_b^3} f'(V/r_b^3) - f(V/r_b^3)\right) - 2\Gamma \pi r_b, \qquad (7)$$

which after non-dimensionalizing by the initial defect length-scale $r_i$ is

$$\mathrm{H}\frac{d\tilde{r}_b}{d\tilde{V}} = \mathrm{M}\frac{\tilde{F}(\tilde{V}/\tilde{r}_b^3)}{\tilde{V}} - \frac{1}{\tilde{V}\tilde{r}_b}, \qquad (8)$$

where $\tilde{\ }$ denotes dimensionless quantities and $\tilde{F} = \frac{V}{r_b^3} f'(V/r_b^3) - f(V/r_b^3)$. There are two key dimensionless variables: the dimensionless friction $\mathrm{H} = \eta g r_i^2 / 4\Gamma$ and the dimensionless elastic modulus $\mathrm{M} = 3\mu r_i / 2\pi\Gamma$. Note that $\tilde{r}_b \geq 1$ since it cannot shrink beyond the initial defect length; therefore, $d\tilde{r}_b/d\tilde{V}$ is constrained to be larger than or equal to 0. We integrate (8) using a forward Euler method and we approximate the contact angle in the model as the unique spherical cap that has the same height $\tilde{z}_{\max}$ at $\tilde{r} = 0$ and same basal intercept $\tilde{r}_b$.

To gain further insight into the model, we interrogate the potential dominant balances in (8). When the volume of the biofilm is small ($\tilde{V} \ll 1$), the lefthand term of (8) is negligible and the righthand two terms balance to give

$$\mathrm{M}\tilde{F}(\tilde{V}/\tilde{r}_b^3)\tilde{r}_b = 1. \tag{9}$$

Since $\tilde{r}_b$ is constrained to be larger than 1, this balance leads to one of two behaviours; *interfacial cavitation* where the gel does not delaminate and $\tilde{r}_b = 1$, or *delamination* where the gel continually detaches from the glass substrate and $\tilde{r}_b > 1$. As the gel delaminates and $\tilde{r}_b$ grows, when $\tilde{V}/\tilde{r}_b^3 \ll 1$, $\tilde{F} \sim \left(\frac{\tilde{V}}{\tilde{r}_b^3}\right)^2$, which means $\tilde{r}_b \sim \tilde{V}^{2/5}$ and the biofilm becomes flatter over time. This scaling behavior is analogous to the scaling behavior observed in "toughness-dominated hydraulic fracture"[18,19] (note that this scaling law is consistent with the initial assumption that $\tilde{V}/\tilde{r}_b^3 \ll 1$ as volume increases). As the biofilm volume and the radius grow, substrate friction becomes relatively important. For most physically relevant parameters, this eventually leads to a balance between friction and elastic deformation such that

$$\frac{H}{M}\frac{d\tilde{r}_b}{d\tilde{V}} = \frac{\tilde{F}(\tilde{V}/\tilde{r}_b^3)}{\tilde{V}}. \tag{10}$$

In this limit, the dynamics depend only on the ratio $H/M = (\pi g r_i/6)(\eta/\mu)$. Although the gel delaminates, the expansion of the delaminating interface is slowed by friction leading to *friction-limited delamination*. As the biofilm grows and the gel delaminates, when $\frac{\tilde{V}}{\tilde{r}_b^3} \gg 1$, $\tilde{F} \sim \left(\frac{\tilde{V}}{\tilde{r}_b^3}\right)^{1/2}$. In this limit, $\tilde{r}_b \sim \tilde{V}^{1/5}$ and the biofilm becomes more dome-like over time (note that this scaling law is consistent with the initial assumption that $\frac{\tilde{V}}{\tilde{r}_b^3} \gg 1$ as volume increases). This regime is analogous to the scaling behavior observed in "viscous-dominated hydraulic fracture," albeit with different scaling laws[18,20]. These three regimes and the transition between them are summarized in Extended Data Figure 5b and c.

We estimated the parameters in the model as follows:

**Growth rate:** The number of bacteria in a given biofilm was measured over time and the growth rate was found to be $g \approx 0.6$ hr.$^{-1}$.

**Gel modulus:** The gel moduli as a function of agarose concentration was measured using shear rheometry and found to be $\log \mu \approx 2.7 \log c + 9.4$ (Extended Data Fig. 1).

**Gel-substrate adhesion and initial defect size:** We fit the contact angle transition both as a function of volume and gel modulus (for mature biofilms) for the non-adhesin-producing mutant. In this case, the mutant exhibits little substrate friction and therefore the contact angle transition is dominated by a balance between interfacial cavitation and delamination. Fitting yields $r_i \approx 5$ μm and $\Gamma \approx 0.02$ N/m[14].

**Friction coefficient:** In Beroz et al.[16] WT* *V. cholerae* bacterial cells were seeded in a microfluidic channel of cross-section $400 \times 60$ µm² and M9 media flowed through the channel at a rate of 500 µL/min. They observed that the cells move at a rate of approximate 2 µm/hr. The drag force on the cells can be approximated as that on a sphere and is $F = 6\pi R \eta_W v_{avg}$ where $R \approx 0.8$ µm is the apparent radius of the cell, $\eta_w \approx 8.9 \times 10^{-4}$ Pa s is the viscosity of water and approximately that of the media, and $v_{avg} \approx 0.35$ m/s is the average fluid velocity in the channel. Taking the footprint of the cell as $A = 3$ µm², the substrate friction coefficient is estimated as $\eta = \frac{F}{A v_{avg}} \approx 2 \times 10^{12}$ Pa s/m.

Limitation of the model: We note that while our model captures most of the features observed in the experiment, it does not predict a bifurcation of contact angle. The predicted change in the contact angle is smooth and no discrete jump in shape or bimodal distribution was predicted. The current course-grained model assumes perfect material properties such as neo-Hookean response of the gel and no stick-slip behavior at the delaminating interface; however, in reality some of these assumptions may not hold. In particular, interfacial fracture (i.e. delamination) is known to be complex and hysteretic[21], which may give additional resistance to the initiation of delamination. The bimodal distribution of contact angles may also be related to intrinsic randomness in adhesin production.

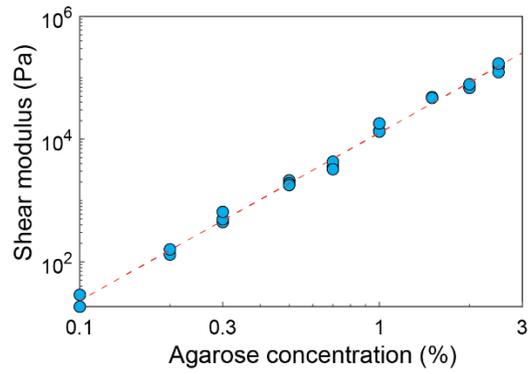

**Extended Data Fig. 1| Dependence of agarose gel stiffness on polymer concentration.** Shear moduli $\mu$ as a function of the agarose concentration $c$. Each data point corresponds to a unique measurement from Zhang et al[11]. The dashed line corresponds to the line of best fit $\log \mu = 2.7 \log c + 9.4$.

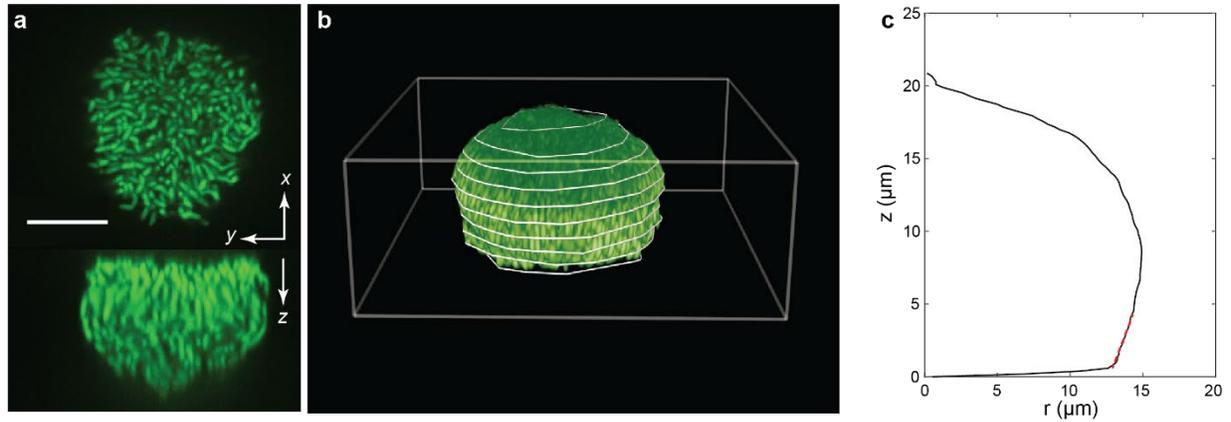

**Extended Data Fig. 2| Example biofilm and contour identification.** (**a**) Raw image showing the basal plane (top) and cross-section (bottom) of a WT* biofilm grown under a 0.5% agarose gel. Scale bar, 10 µm. (**b**) Three-dimensional reconstruction of the biofilm in (a) with the areal convex hulls overlain (white). (**c**) Effective radii of the convex hulls as a function of the height of the biofilm. Red line corresponds to a linear fit from which the effective contact angle is calculated.

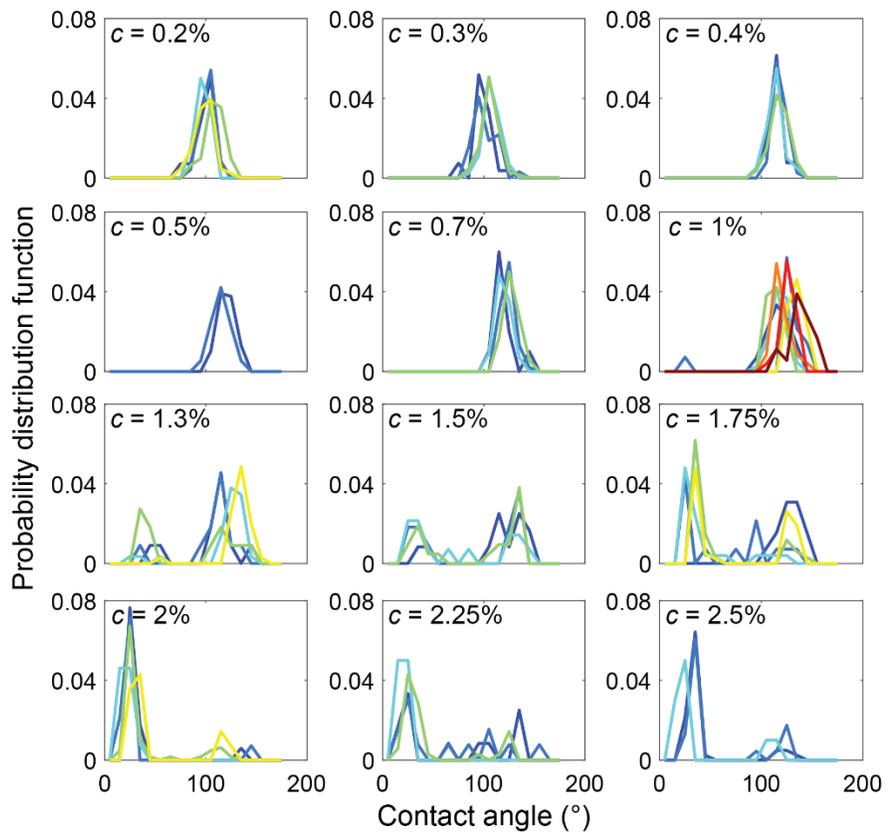

**Extended Data Fig. 3| Contact angle distributions across experiments.** Probability distribution function of different contact angles for biofilms grown under different stiffness gels. Each line corresponds to a distinct single field of view with at least 10 biofilms. In general, we find that the distributions, including the bimodal distributions at intermediate concentrations, are well preserved across experiments.

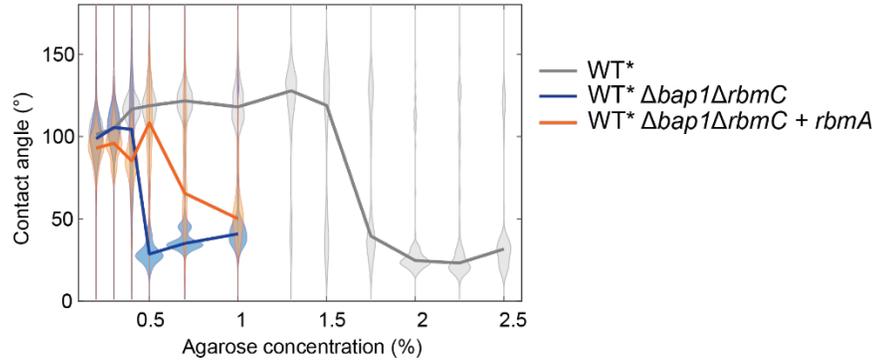

**Extended Data Fig. 4| Contact angle distributions across mutants.** Violin plot of contact angles calculated for biofilms formed by different mutant strains grown under gels of different agarose concentrations. Each chord represents a probability distribution and the lines connect the median values of the distributions. The grey data corresponds to the data in Fig. 1c, the blue data is for a mutant strain that lacks biofilm adhesins Bap1 and RbmC and the orange data is for a mutant strain that also lacks biofilm adhesins Bap1 and RbmC but expresses the cell-to-cell adhesin RbmA. We note that the cell-to-cell adhesion seems to minimally affect the shape transition.

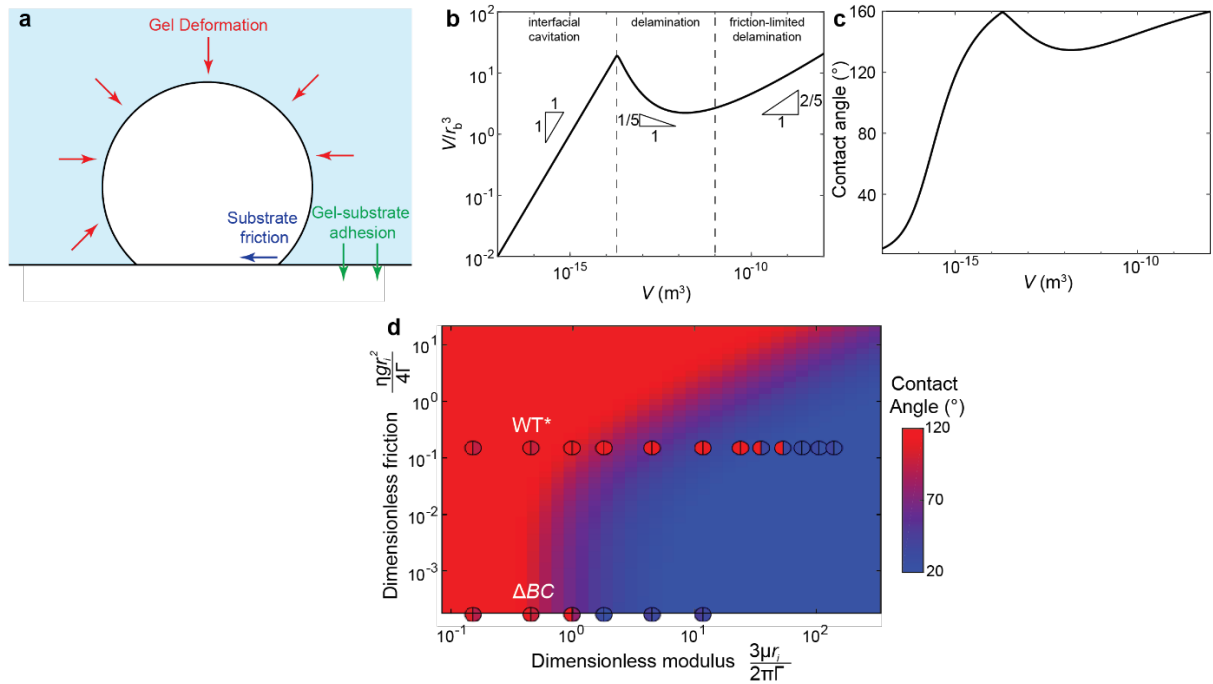

**Extended Data Fig. 5| Competition between gel stiffness and substrate friction controls biofilm morphogenesis.** (**a**) Schematic of the theoretical setup. A biofilm with basal radius $r_b$ sits at the interface of a rigid bottom substrate and a semi-infinite elastic gel (blue). As the biofilm grows its expansion is impeded by friction from the substrate and growth of the biofilm deforms the gel around it potentially delaminating the gel from the substrate. (**b**, **c**) Example solutions showing the evolution of the rescaled volume $V/r_b^3$ (b) and contact angle (c) for $\mu = 3$ kPa and $\eta = 10^{11}$ Pa s/m. Experimentally, the initial regimes are difficult to observe because of errors in defining the shape of a biofilm consisting of tens of cells. (**d**) Predicted biofilm contact angle as a function of dimensionless substrate friction and gel modulus. Overlain circles denote the experimental results from Extended Data Fig. 4. The two halves of each circle quantify the interquartile range of measured contact angles. The adhesin-less mutant $\Delta bap1 \Delta rbmC$ ($\Delta BC$) has a negligible dimensionless friction value and is therefore plotted on the x-axis.

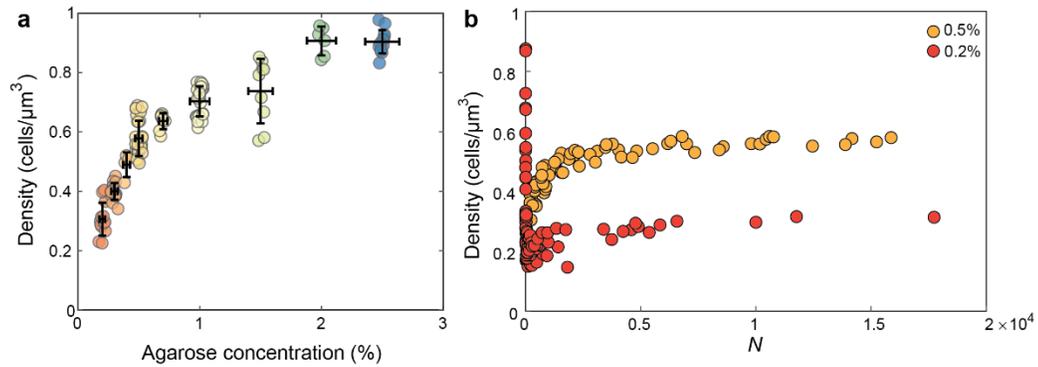

**Extended Data Fig. 6| Cell density as a function of agarose concentration and time. (a)** Cell density of mature biofilms for different agarose concentrations. Vertical error bars correspond to s.d. in measurements and horizontal error bars correspond to the uncertainty in agarose concentration. **(b)** Cell density as a function of the number of cells for 4 and 8 biofilms measured over time at 0.2% and 0.5% agarose concentrations respectively. In both cases, the density tends to a plateau instead of continually increasing which suggests that the confining pressure tends to a constant rather than increasing unboundedly. This observation is consistent with the energetic model where the pressure in dome-shaped biofilms is predicted to tend to a constant value with increasing volume.

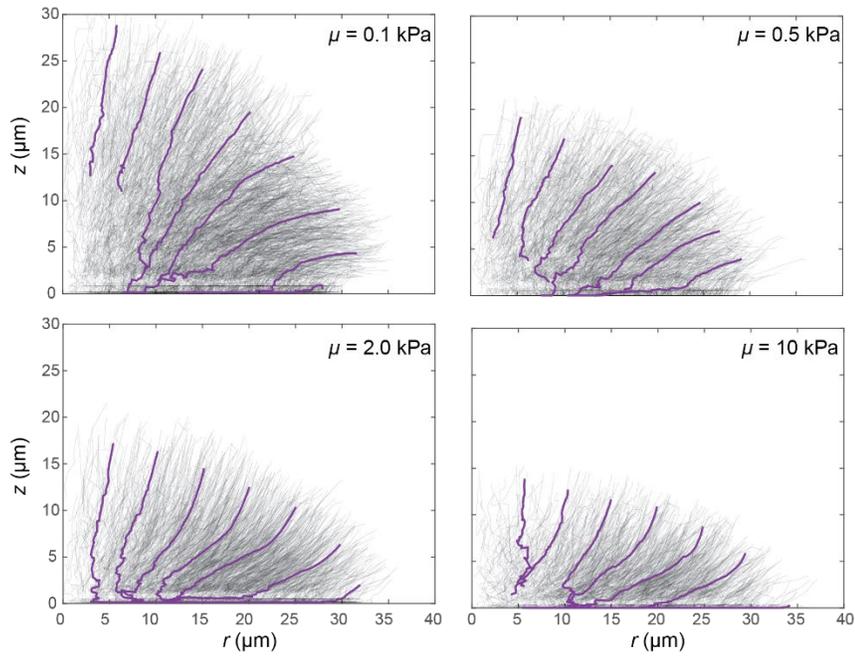

**Extended Data Fig. 7| Cell trajectories in agent-based simulations also show a pattern in response to gel stiffness.** Trajectories of cells in agent-based simulations with different gel stiffnesses show two different types of patterns; either curving down leading to fountain-like trajectories (top) or curving up (bottom), consistent with the experimental observation.

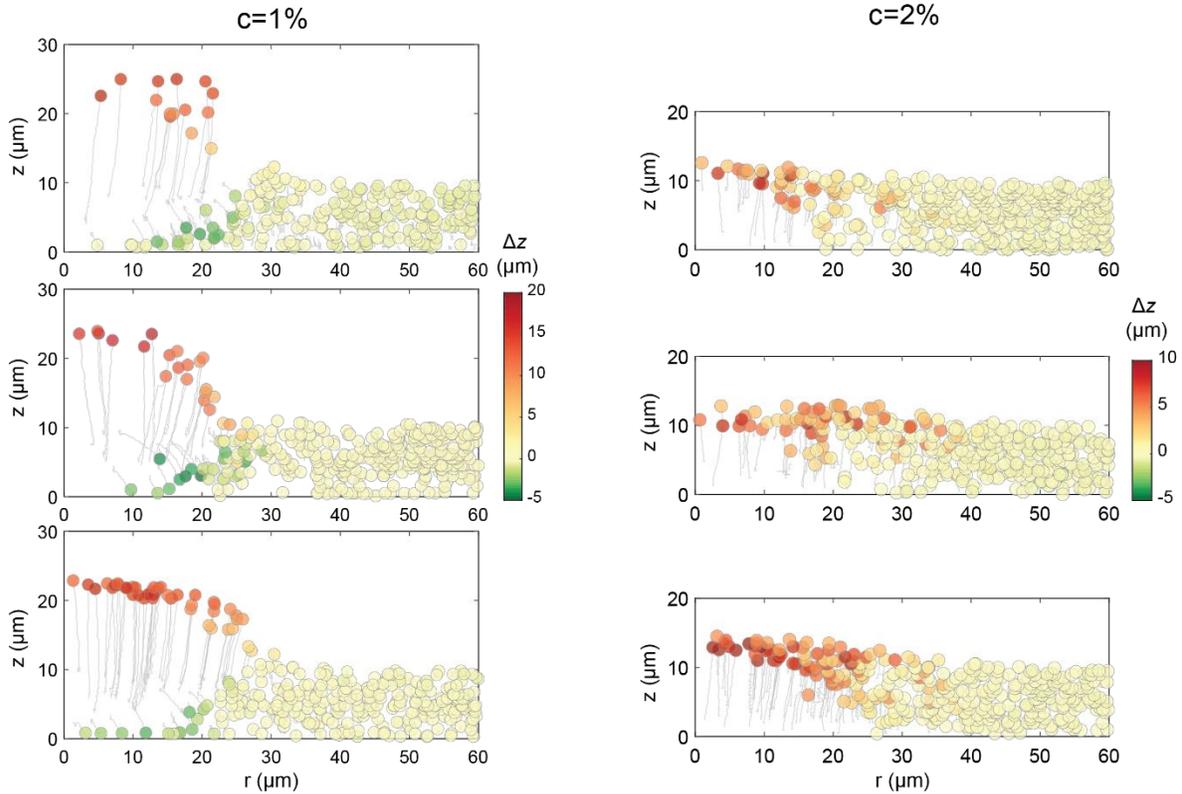

**Extended Data Fig. 8| Distinct gel deformation modes for dome- and lens-shaped phenotypes.** Displacement of tracer particles in the axisymmetric coordinates of the biofilm during growth of 6 different biofilms. The colors denote the direction and magnitude of the vertical displacement of the beads at the end of the experiment with respect to the original location $(z(t) - z(0))$. Consistent with the interfacial cavitation model for the growth of dome-shaped biofilms, we observed negative values near the boundary, corresponding to gel materials that are compressed and therefore move closer to the glass substrate.

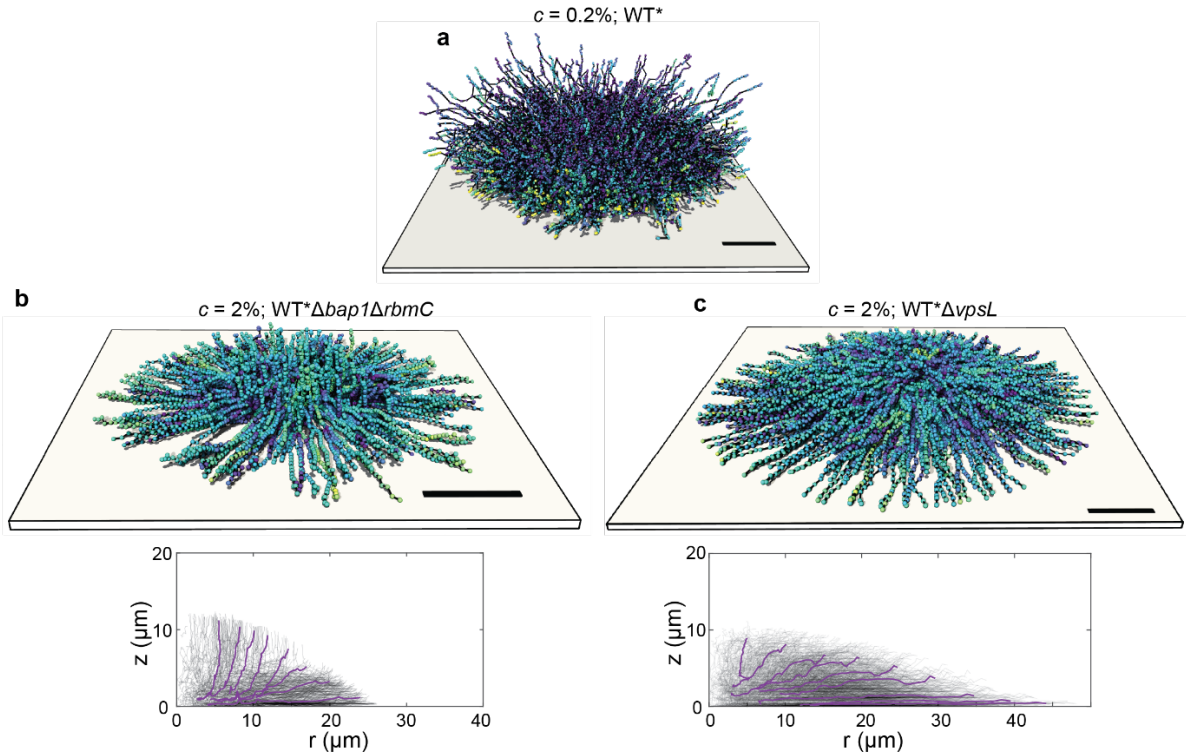

**Extended Data Fig. 9| Cell trajectories in mutant biofilms. (a)** Reconstructed puncta trajectories for a WT* biofilm grown under a soft gel (corresponding to averaged data in Fig. 3b). (**b**, **c**) 3D reconstructed puncta trajectories (top) and projected and averaged trajectories (bottom) for a biofilm that does not produce the extracellular adhesins Bap1 and RbmC (b) and for bacteria that do not produce any extracellular matrix (Δ*vpsL*, c) grown in a stiff gel ($c = 2\%$). While the Δ*bap1*Δ*rbmC* mutant (b) follows similar trajectories as the WT* biofilm in a stiff environment (Fig. 3b), trajectories of Δ*vpsL* cells exhibit the opposite curvature. It has been shown previously that the Δ*bap1*Δ*rbmC* mutant still retains some adhesion to the top gel surface through the exopolysaccharide[11], which is critical to create the upward bending of the cell trajectory. In contrast, the Δ*vpsL* shows a trajectory that can be expected if all cells are growing in dimensions proportional to the growing radius and height. These results support the conclusion that biofilm shape and biofilm-gel adhesion jointly dictate the cell trajectories in a biofilm.

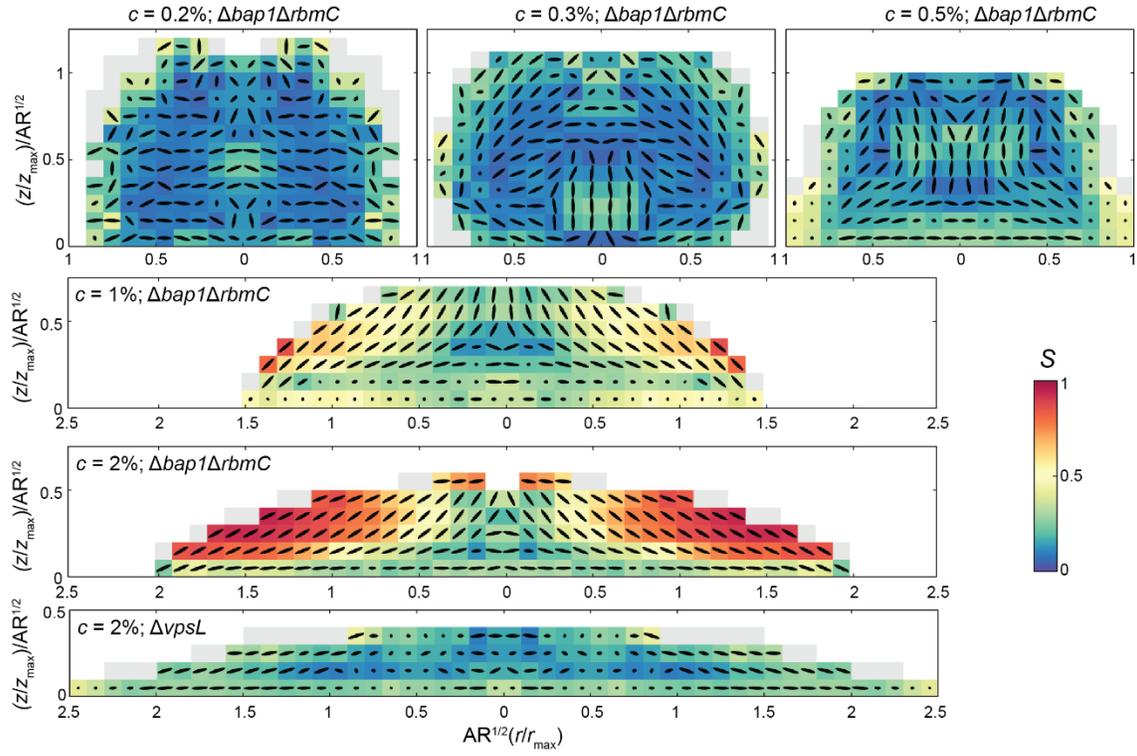

**Extended Data Fig. 10| Cell organization in mutant biofilms.** Azimuthally averaged cell orientations for mutant biofilms grown in different stiffness environments. The averaging is performed in the same manner as Fig. 5. Colors denote the nematic order parameter (degree of order) and the ovals denote the average director of the cells projected into $(r, z)$ space. A total of $\{7, 8, 4, 6, 7\}$ biofilms were averaged for the $\Delta bap1\Delta rbmC$ mutant with $c = \{0.2\%, 0.3\%, 0.5\%, 1\%, 2\%\}$ respectively, and 6 biofilms were averaged for the $\Delta vpsL$ mutant with $c = 2\%$.

**Table S1: List of the strains used in this study**

| Strains | Genotype | Source |
|---|---|---|
| JN007 | $vpvc^{W240R}$, $\Delta vc1807$::Ptac-mNeonGreen-Spec$^R$ | 4 |
| JN008 | $vpvc^{W240R}$, $\Delta rbmA$, $\Delta vc1807$::Ptac-mNeonGreen-Spec$^R$ | 4 |
| JN010 | $vpvc^{W240R}$, $\Delta rbmA$, $\Delta bap1$, $\Delta rbmC$, $\Delta vc1807$::Ptac-mNeonGreen-Spec$^R$ | 4 |
| JN011 | $vpvc^{W240R}$, $\Delta vpsL$, $\Delta vc1807$::Ptac-mNeonGreen-Spec$^R$ | 4 |
| JN036 | $vpvc^{W240R}$, $\Delta rbmA$, $\Delta bap1$, $\Delta rbmC$, $\Delta vc1807$::Ptac-mNeonGreen-Spec$^R$, pJY057 | 4 |
| JN148 | $vpvc^{W240R}$, $\Delta rbmA$, $\Delta lacZ$::Ptac-mNeonGreen-µNS, $\Delta vc1807$::Ptac-mScarletI-Spec$^R$ | 4 |
| JN150 | $vpvc^{W240R}$, $\Delta rbmA$, $\Delta bap1$, $\Delta rbmC$, $\Delta lacZ$::Ptac-mNeonGreen-µNS, $\Delta vc1807$::Ptac-mScarletI-Spec$^R$ | 4 |
| JN151 | $vpvc^{W240R}$, $\Delta vpsL$, $\Delta lacZ$::Ptac-mNeonGreen-µNS, $\Delta vc1807$::Ptac-mScarletI-Spec$^R$ | This study |
| | | |
| **Plasmid** | | |
| pJY057 | Kan$^R$, $araC$-$P_{BAD}$-$bap1$ | 22 |